# Moving magnetic domain walls with sound alone

Alejandro Rivelles[1,2*], Rocío Yanes[3,4*], Luis Torres[3,4], Manuel Abuín[1,2], Javier Grandal[1], Maedeh Sepehr[1], Guzmán Orero-Gámez[5], Rodrigo Guedas[1,6], Laura Fernández-García[7], Raúl Izquierdo-López[1,8], Marco Maicas[1], Maria del Mar Sanz[1], Jorge Pedrós[1,8], Fernando Calle[1,8], Sandra Ruiz-Gómez[9], Muhammad Waqas Khaliq[9], Miguel Angel Niño[9], Saül Vélez[5,10], Michael Foerster[9], Luis López-Díaz[3,4♦], Jose Luis Prieto[1,♦]

[1] Instituto de Sistemas Optoelectrónicos y Microtecnología (ISOM), Universidad Politécnica de Madrid, Madrid, Spain.
[2] Department of Electrical Engineering, Automatic Electronics and Applied Physics, ETSIDI, Universidad Politécnica de Madrid, Madrid, Spain.
[3] Applied Physics Department, Universidad de Salamanca, Salamanca, Castilla y León, Spain.
[4] Unidad de Excelencia en Luz y Materia Estructuradas (LUMES). Universidad de Salamanca. Salamanca, Spain.
[5] Spintronics and Nanodevices Laboratory, Departamento de Física de la Materia Condensada and Condensed Matter Physics Center (IFIMAC), Universidad Autónoma de Madrid, Spain
[6] Université Grenoble Alpes, CNRS, CEA, Grenoble INP, SPINTEC, 38000 Grenoble, France.
[7] Departamento de Física de Materiales, Universidad Complutense de Madrid, Madrid, 28040, Spain.
[8] Departamento de Ingeniería Electrónica, ETSI Telecomunicación, Universidad Politécnica de Madrid, Madrid, Spain.
[9] ALBA Synchrotron Light Facility, Carrer de la Llum, 2 – 26, Cerdañola del Valles, 08290 Barcelona, Spain.
[10] Instituto Nicolás Cabrera, Universidad Autónoma de Madrid, Spain

[*,♦] These authors contributed equally.

## Abstract

Surface Acoustic Waves (SAW) have been used in spintronic applications to decrease the magnetic field or the electric current required to act on the magnetization. A common belief is that a SAW alone cannot achieve a directed magnetic switching in a device without an assisting magnetic field or electric current. In this work, we demonstrate magnetic domain wall motion driven solely by an acoustic wave. Using XMCD-PEEM, we show extensive evidence of SAW-induced and field-free magnetic domain wall motion (DW) in the direction of the wave propagation. Our micromagnetic simulations reveal a mechanism that allows the SAW to transfer linear momentum to the DW. Experimentally, the largest DW average velocity measured was ~12 m/s, although our simulations predict that velocities in the range of 100 m/s could be attained. This new mechanism opens the door to designing innovative spintronic devices where the magnetization can be controlled exclusively by an acoustic wave.

## Introduction

Magnetostriction is a property of magnetic materials through which their magnetic properties can be altered using mechanical stress. This property is at the core of several widespread applications such as underwater sonar[1], security labels[2], and magnetic sensors[3]. The field of spintronics has also explored the possibilities of magnetoelastic interactions through numerous experiments, often using a Surface Acoustic Wave (SAW)[4,5,6]. Weiler *et al.*[7] showed that an elastic wave could lead to an acoustic-driven ferromagnetic resonance (FMR) in a Nickel film. The same authors also produced a pure spin current and measured magnetoelastic spin pumping[8]. The SAW can assist the magnetic switching of a ferromagnetic film[9] or of a nano-element[10,11], it can reduce the energy required for magnetic recording[12] or even decrease the energy dissipation in spin-transfer-torque random access memories[13]. More recently, a SAW was used to achieve a controlled generation of magnetic skyrmions[14] or to promote field-free switching in local regions of a magnetic semiconductor[15]. In spintronic devices that rely on controlling the magnetic domain wall (DW) motion, the interaction with a SAW can facilitate the DW motion and decrease the electric current[16] required for operation, which often leads to detrimental thermal effects[17,18]. There is also a theoretical work proposing the use of an acoustic standing wave to control the movement and position of a DW[19].

Despite the considerable effort devoted by the scientific community to explore the interaction of a SAW with magnetic devices, the goal of reliably moving a DW in the direction of the SAW propagation, without any external magnetic field or electric current, has not been achieved. As the SAW is a harmonic oscillation, one should not expect a net unidirectional DW movement: the DW would move forward during half of the SAW cycle and, the same distance backwards during the other half. Very recently, Yang et al.[20] reported that, in the presence of an out-of-plane magnetic field, a Skyrmion could be occasionally moved under the action of a shear horizontal acoustic wave. At best, the Skyrmion would move 37% of the attempts in small steps equivalent to speeds in the range of μm/s.

Here we show that a SAW can move a DW in the same direction as the wave propagation. Using XMCD-PEEM, we show evidence that the SAW moves the DWs in ferromagnetic nano-strips with zero external magnetic field, always in the direction of the SAW propagation. The micromagnetic simulations unveil a mechanism that allows the SAW (an acoustic wave) to transfer linear momentum to the DW (a magnetic texture). During half of the SAW cycle, the asymmetric transversal DW reduces its width by accumulating magnetoelastic energy. When this energy becomes sufficiently large, the DW jumps forward to release some of the magnetoelastic stress through an internal fluctuation, resulting in a net movement in the direction of the SAW. In this experiment, the largest DW average velocity observed was ~12 m/s, although the micromagnetic simulations predict that velocities larger than 100 m/s could be attained in realistic experimental conditions.

### Results

The device configuration used for the main experiment is sketched in Fig. 1a. The Interdigital Transducers (IDT) are deposited over a 128º Y-cut $LiNbO_3$ substrate, designed to resonate at 1.3 GHz for the Rayleigh mode. The characterization of the IDT can be found in the Suppl. Info. S1. A collection of strips with widths ranging from 500 nm to 1.6 μm, each incorporating an elliptical nucleation pad, are situated 3 mm away from the IDTs, as shown in Fig. 1a. The large distance between the IDT and the magnetic devices was required to avoid possible arcs on the sample from the imaging extraction voltage of the XMCD-PEEM set-up. The structure of the magnetic strip is Cr(1nm)/FeCo(2nm)/FeCoB(3 nm)/Py(1nm)/Pt(1nm), where FeCo stands for $Fe_{65}Co_{35}$, FeCoB for $(Fe_{65}Co_{35})_{95}B_5$ and Py for Permalloy $Ni_{80}Fe_{20}$. The FeCo layer is magnetostrictive, and it was the material that, via resistive measurements, gave us the first evidence of the mechanism reported in this work. FeCoB is also magnetostrictive and helps reduce the coercivity as shown in the hysteresis loops in Suppl. Info. S2. Finally, the thin Py layer on top reduces the coercivity even more and decreases the probability of DW pinning in the strips.

Figure 1 shows sequences of XMCD-PEEM images acquired on 900 nm-wide strips, measured at the Fe $L_3$ edge (see Methods for details). The procedure to nucleate and inject a DW in the strip is written on top of the first sequence of images. First, the strip is saturated in plane with a maximum field of 200 Oe. Then, an opposite field of 45 Oe is applied, reversing the magnetization in the nucleation pad. Finally, the field is ramped to zero and stays at zero while the experiment is performed by applying SAW pulses. In Fig. 1b, starting from saturation to the right, once the field sequence is completed and the field is brought to zero, the XMCD-PEEM image reveals a single Tail-to-Tail DW close to the injection pad. Then, a 500 µs 25 dBm SAW pulse is delivered, still at zero external magnetic field. The image taken after the pulse (Fig. 1c) shows a clear displacement of the DW in the same direction as the SAW propagation, from left to right. A second SAW pulse completes the reversal (Fig. 1d), implying that the DW was displaced again in the same direction as the SAW propagation at zero field. The situation is very similar when a Head-to-Head DW was injected, as shown in the sequence Fig. 1e to Fig. 1g. Although in Fig.1 the Head-to-Head DW seems less mobile than the Tail-to-Tail, in all the sequences measured, we could not find any conclusive difference in the response of the two types of DW to the acoustic wave. In the Supplementary Information, we provide additional experimental evidence of the particularities of the SAW-driven DW displacement at zero external magnetic field.  Figure S4 shows examples of the movement of the two DW types under the sole action of the SAW. Figure S5 demonstrates that the domain wall (DW) moves in successive steps following four identical consecutive SAW pulses. Figure S6 shows that the DW displacement increases with SAW pulses of greater width, although some stochasticity is observed in the pinning and motion of the DW. Figure S7 reveals that DW displacement also increases with the amplitude of the SAW pulses, with no displacement observed below 25 dBm. Figure S8 provides evidence that the DW can be displaced solely by the SAW when the acoustic wave propagates from the tip of the strip toward the nucleation pad. Remarkably, the SAW alone can reverse the magnetization of the strip, even in the elliptical pad, which is significantly wider than the strip. Finally, in Figure S9, we show images demonstrating that the SAW can move the DWs in the direction of the wave propagation (left to right), even against the opposing pressure of a small applied magnetic field.

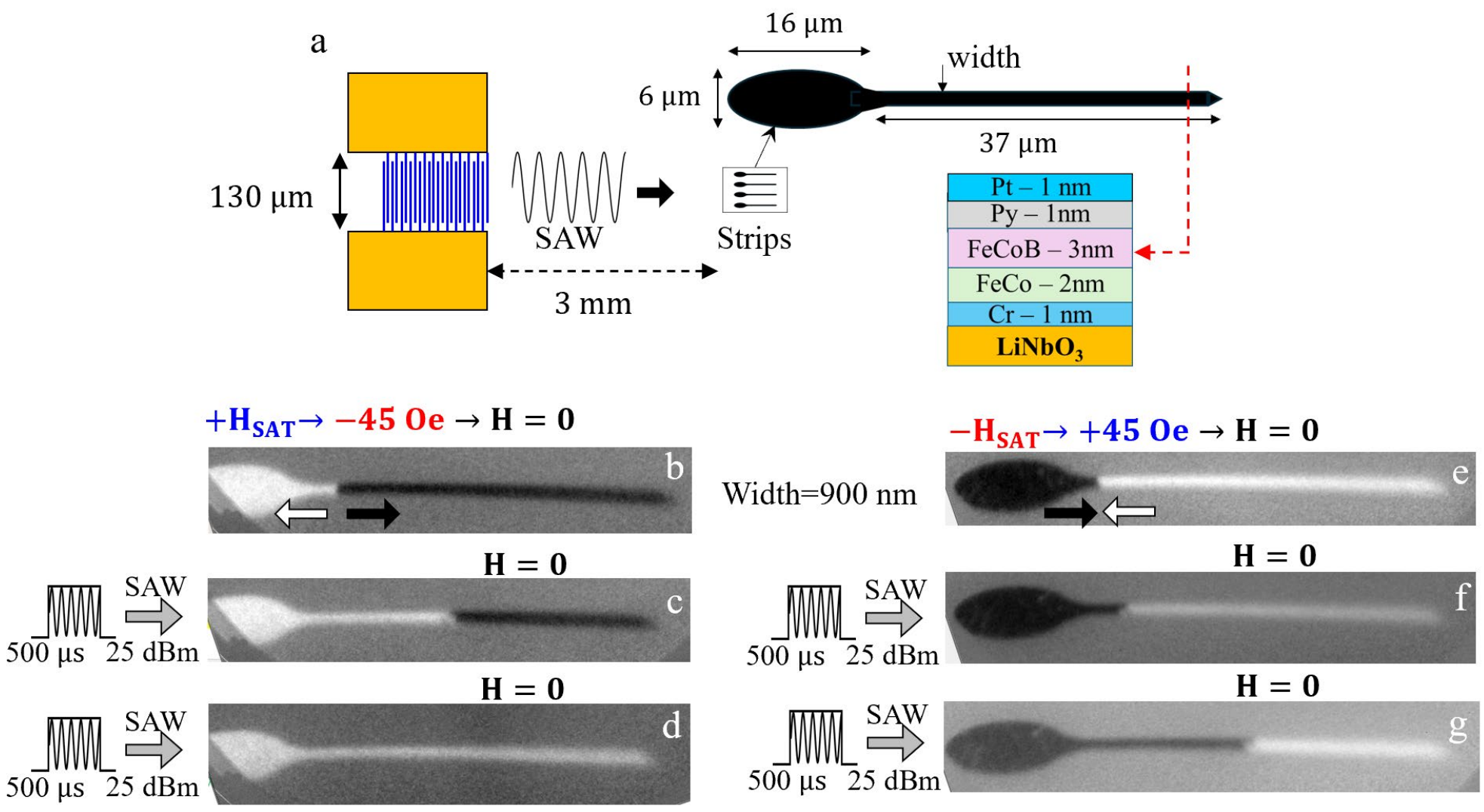

**Fig.1**. Experiment description. (a) Schematic representation of the sample. An IDT, deposited over $LiNbO_3$, is placed 3 mm away from a set of ferromagnetic strips of different widths. The strips have an elliptical nucleation pad with the dimensions shown in the figure. The length of the strip is always 37 µm. In this figure, all the strips are 900 nm wide. (b) XMCD-PEEM image taken at zero external magnetic field, after the field sequence shown on top of the figure. The arrows indicate the direction of the magnetization. (c) XMCD-PEEM image taken after a 500 µs 25 dBm SAW pulse, always at zero external magnetic field (d) XMCD-PEEM image taken after a second 500 µs 25 dBm SAW pulse, also at zero external magnetic field. (e-g) Same sequence as (b-d) but with a Head-to-Head DW instead.

In all the extensive evidence obtained by XMCD-PEEM microscopy, the movement of the DW was always observed to be in the direction of the SAW propagation (left to right).  To rule out any major role played by the heat carried by the SAW, we used a thermography camera to measure the temperature in the sample. The details can be found in Suppl. Info. S4. At 28 dBm of continuous SAW, the thermal increase in the devices (3 mm away from the IDT) is less than 10 ºC in steady state. The temperature in the IDT is larger, but it does not reach 20 ºC at the maximum power used in this experiment. This small temperature increase agrees with previous estimations by other authors[21,22] and, although it may be relevant for materials with perpendicular magnetic anisotropy[21], it is small for materials like the ones used in this work, where the DWs do not move by thermal activation. Note also that the experiments shown in this work use a pulsed SAW, so the temperature increase in the sample is most likely smaller than 10ºC. Finally, we ought to mention that the temperature gradient along the entire length of the ferromagnetic strip is smaller than 0.04 ºC/µm. This small thermal gradient would unlikely play any role in the DW movement, which would be towards the hot end, i.e., against the direction of the wave propagation and the DW motion observed.

Figure 2 shows the influence of the strip width on the zero-field DW motion driven by the SAW. The sequence of XMCD-PEEM images in Figs. 2a-c, taken for four different strip widths and using a short 2 µs SAW pulse, show that strips 1.2 µm and 1 µm wide allow a larger displacement of the DW than wider or narrower strips. Figure 2d summarizes the DW velocity in the same four strips as a function of the SAW power for a SAW pulse 2 µs long. The velocity is measured by dividing the DW displacement by the 2 µs of the SAW pulse. The minimum pulse width we could apply during the Synchrotron experiment was 2 µs. Moreover, given the length of the strips (which had to allow the entire strip to be visible within the 50 µm field of view of the XMCD-PEEM microscope), it was not possible to perform a detailed study of the DW velocity for longer pulse widths.

Although Figure 2d shows a clear increase in velocity with increasing SAW amplitude (see also Figure S7 in the Supplementary Information), there is some variability in the results. In most of the experimental images presented in this work, it is evident that DW displacement exhibits a degree of stochastic behaviour, possibly due to the polycrystalline nature of the ferromagnetic strip and lithographic imperfections. This stochastic behaviour is also predicted by our micromagnetic simulations discussed later.  Despite the dispersion in the results, strips 1 µm and 1.2 µm wide show, on average, a larger DW velocity for almost any SAW power. The maximum velocity measured was ~12 m/s. To image the type of DW present in the strips, we used Magnetic Force Microscopy (MFM). In all the strips under study, we only observed transversal domain walls, such as those shown in Fig. 2e for strips 1.2 µm wide and in Fig. 2f for strips 1 µm wide. The shape and width of the transversal DWs in wider or narrower strips may vary slightly from those shown in Figs. 2e-f.

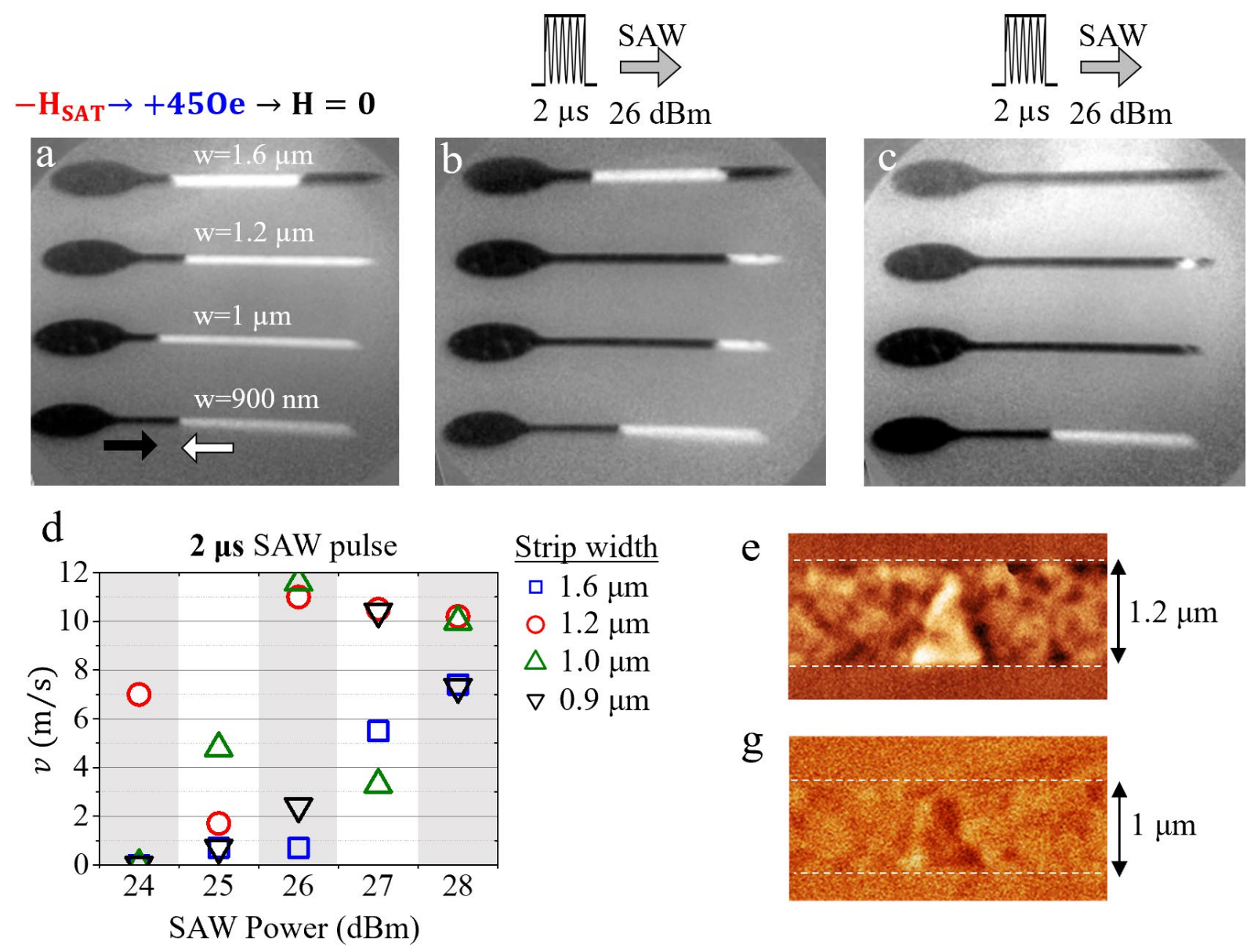

**Fig. 2.** Magnetic domain wall movement under the action of a SAW in strips of different widths. (a) XMCD-PEEM image of the magnetic configuration of four strips of various widths, as marked in the figure, after the field sequence shown on top of the figure.

The arrows indicate the direction of magnetization. (b) Magnetic state of the same strips in (a) after a 2 µs 26 dBm SAW pulse and zero external magnetic field. (c) Magnetic state of the same strips in (b) after a second 2 µs 26 dBm SAW pulse and zero external magnetic field. (d) Plot with the DW velocity versus the SAW power, measured in the four strips shown in (a-c) after applying a single 2 µs SAW pulse with zero field (e) MFM image of the DW present in strips 1.2 µm wide (f) MFM image of the DW present in strips 1 µm wide.

The evidence shown in Fig. 2 implies that the field-free DW motion in the direction of the SAW propagation also depends on the width of the strip. We have conducted a detailed theoretical study with micromagnetic simulations to clarify the mechanism that allows the DW to move under the sole action of the SAW and only in the direction of the wave propagation.

The micromagnetic simulations were performed with Mumax 3 software[23], where the action of the SAW was included via the magnetoelastic field as explained in Methods. The simulation shown in this manuscript involves only one layer of the experimental stack—the FeCo layer—for two main reasons. Firstly, adding additional magnetic layers significantly increases the computational cost. Secondly, in preliminary experiments using strips made of a single FeCo layer, we observed DW motion in the direction of the wave propagation with a small external magnetic field present. Indeed, it will become clear that the simulation with strips made of a single FeCo layer accounts for all the experimental findings.

In Fig. 3a we show a schematic representation of the strip simulated, a single layer of FeCo, 3 nm thick and 1 µm wide. The SAW can be delivered from left to right, called SAW+, and from right to left, called SAW-. As shown in Fig. 3a, an asymmetric transverse DW like the ones observed by MFM is stable in the strip. The mechanical excitation is activated after the DW is stabilized at the center of the strip. Fig. 3b displays the time evolution of the position along the nanostrip, $X_{DW}$, for two values of the SAW amplitude $\varepsilon$ and both propagation directions, left to right or SAW+ and right to left or SAW-. The equivalence between the SAW amplitude used in the simulations (in ppm) and the SAW amplitude used in the experiments (in dBm) can be found in Suppl. Info. S1. Without SAW (black line), the DW remains at the centre ($X_{DW} = 0$) but if the SAW is activated, the DW undergoes a complex oscillatory motion that results in a net displacement along the direction of the SAW propagation. The process can be visualized in the animation in Suppl. Mat. (Movie 1). In Fig. 3c, we present snapshots of the magnetization configuration for different values of the amplitude of the SAW, taken at the same instant ($t = 15$ ns), showing that the DW moves along the direction in which the SAW propagates, and that the displacement increases with the amplitude of the SAW.

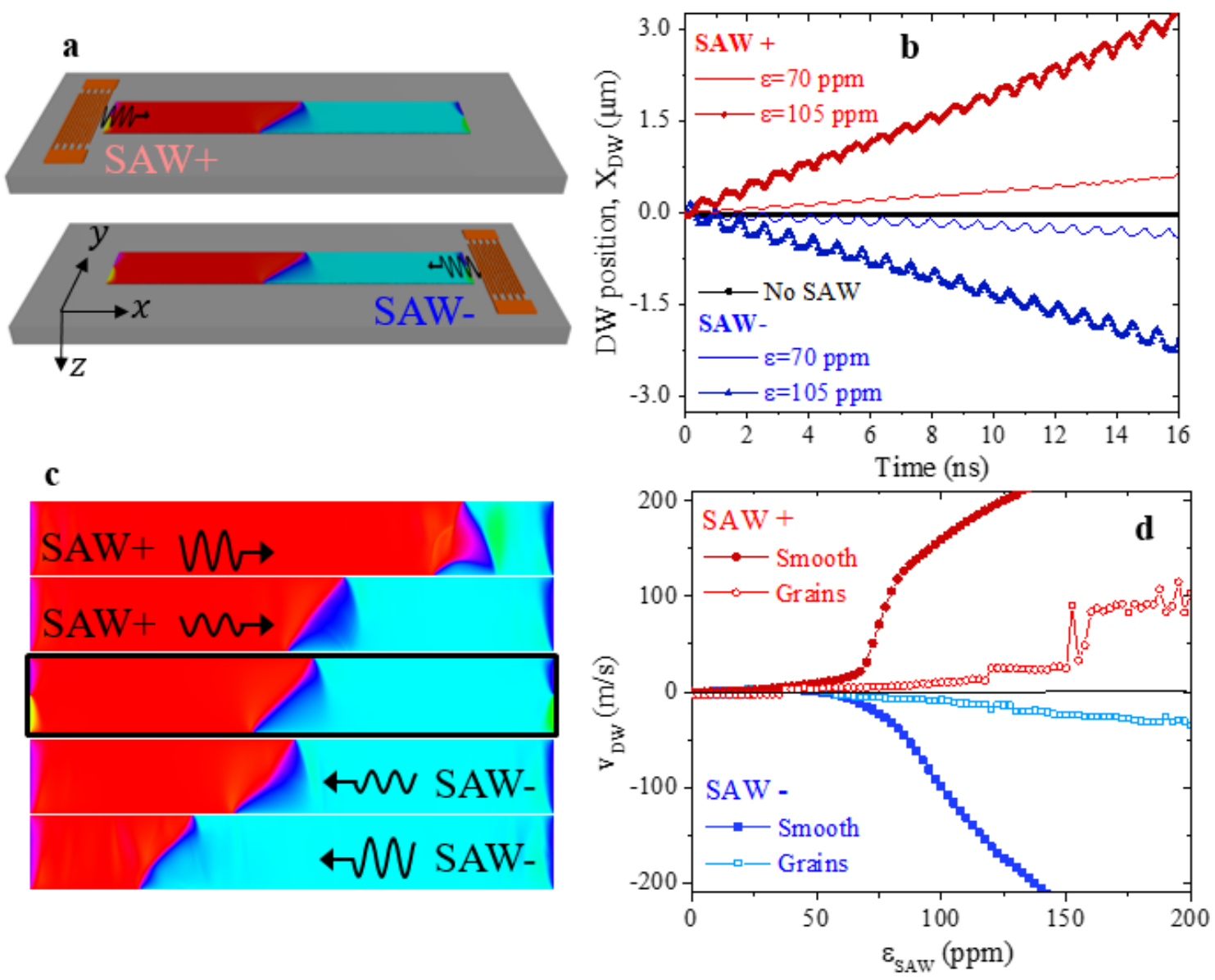

**Fig. 3.** Micromagnetic simulations of the acoustic movement of a magnetic domain wall. (a) Set up with a single transversal DW nucleated in the middle of a FeCo strip where the SAW is delivered from both directions (b) Time evolution of the DW position along the nanostrip for several values of the SAW amplitude and propagation direction. (c) Snapshots of the magnetization configuration 15 ns after activating the SAW, where the color scale represents the magnitude of $m_x$ (red for $m_x = 1$, cyan for $m_x = -1$), for SAW

amplitudes of 70 ppm and 105 ppm and opposite propagation directions. (d) DW average velocity versus the amplitude of the SAW for ideal-smooth and granular nanostrips in both SAW+ and SAW- directions.

In Fig. 3d, the average DW velocity, $v_{DW}$, is plotted as a function of the SAW amplitude, $\varepsilon_{SAW}$, for both SAW propagation directions. The DW speed increases monotonically with the amplitude of the SAW, and three regimes can be observed. After a linear regime with a small slope for low values of $\varepsilon_{SAW}$, a rapid non-linear increase is found for intermediate values ($\varepsilon_{SAW} \sim 70 - 100$ ppm) followed by a recovery of the linear regime ($\varepsilon_{SAW} > 100$ ppm) with a larger slope than in the initial regime. Together with the velocity plots obtained for ideal samples (solid symbols), in Fig. 3d we also show the result obtained for samples with disorder (open symbols). Specifically, the disordered samples have a grain distribution of characteristic diameter $d = 20$ nm and a standard deviation of the anisotropy constant among grains $\sigma = 3\%$. The DW velocity reduces significantly due to the pinning introduced by the disorder, and the regimes mentioned above are blurred. Nevertheless, the main effect remains: the SAW pushes the DW in the direction of its propagation. In Figs. 3b and 3d, we also observe that the velocity for SAW+ and SAW- is different. This is due to the asymmetric configuration of the transversal DW, which is tilted with the $x$ axis (see snapshots in Fig. 3c), resulting in an asymmetric response for opposite SAW propagation directions. This effect is highlighted in Fig. S11 of Suppl. Inf., where the DW speed is plotted as a function of the angle between the nanostrip axis and the SAW propagation direction for the four possible DW configurations with equivalent energy. The maximum average velocity for each type of DW is obtained for a given angle, showing that the SAW-induced DW motion reported in this work distinguishes between different DW tilt and chirality configurations. This fact will become clearer after the discussion section. We should also mention that the longitudinal strain wave $\varepsilon_{xx}$ is the responsible for the zero-field DW motion[24,25], as shown in Suppl. Inf. S6. Additionally, the movement of the DW can be enhanced by reducing the magnetic damping, as shown in Suppl. Inf. S7.

## Discussion

In this section, we provide an intuitive explanation of the observed mechanism that allows a field-free movement of the DW in the direction of the SAW propagation. In Figure 4a-e we plot the time evolution of the energy of the DW, normalized by its energy without SAW, $\sigma_{DW}/\sigma_{DW}^{0}$ (Fig. 4a), the width of the DW, $\delta_{DW}$ (Fig. 4b), its length, $\Delta L_{DW}$ (Fig. 4c), its instantaneous velocity, $v_{DW}$ (Fig. 4d) and the DW position, $X_{DW}$ (Fig. 4e). In all the panels, for clarity, we plot the longitudinal strain of the SAW $\varepsilon_{xx}$ (dash black line) evaluated at the center of the moving DW. In all these panels, there is a yellow band around 1 ns, which marks the region where we focus this discussion.

We begin by noticing that the energy of the DW becomes maximum just after the SAW reaches its maximum stress (Fig. 4a), which coincides with the minimum in its width (Fig. 4b), i.e. the DW is compressed. At that point, the DW finds a strategy to liberate part of this accumulated energy: it jumps slightly forwards (see Fig. 4e) and that allows the DW to increase its size again (Fig. 4b and 4c) and reduce its energy faster than during its normal oscillatory behaviour (Fig. 4a at the yellow band, just after the maximum). Although at this specific time, the instantaneous velocity of the DW changes sign momentarily, the overall effect on the instantaneous velocity (Fig. 4d) is that, during one SAW cycle, the DW spends slightly more time with positive velocity than with negative velocity, resulting in a net displacement in the direction of the SAW propagation. The mechanism can be visualized during several cycles of the SAW in Movies 1, 2 and 3 in Suppl. Mat., where it is visible how the DW releases part of its accumulated magnetoelastic energy through an internal wave that travels along the length of the DW as the DW 'springs' slightly forward. The overall result is that the DW gains linear momentum, transferred from the SAW by this non-trivial mechanism.

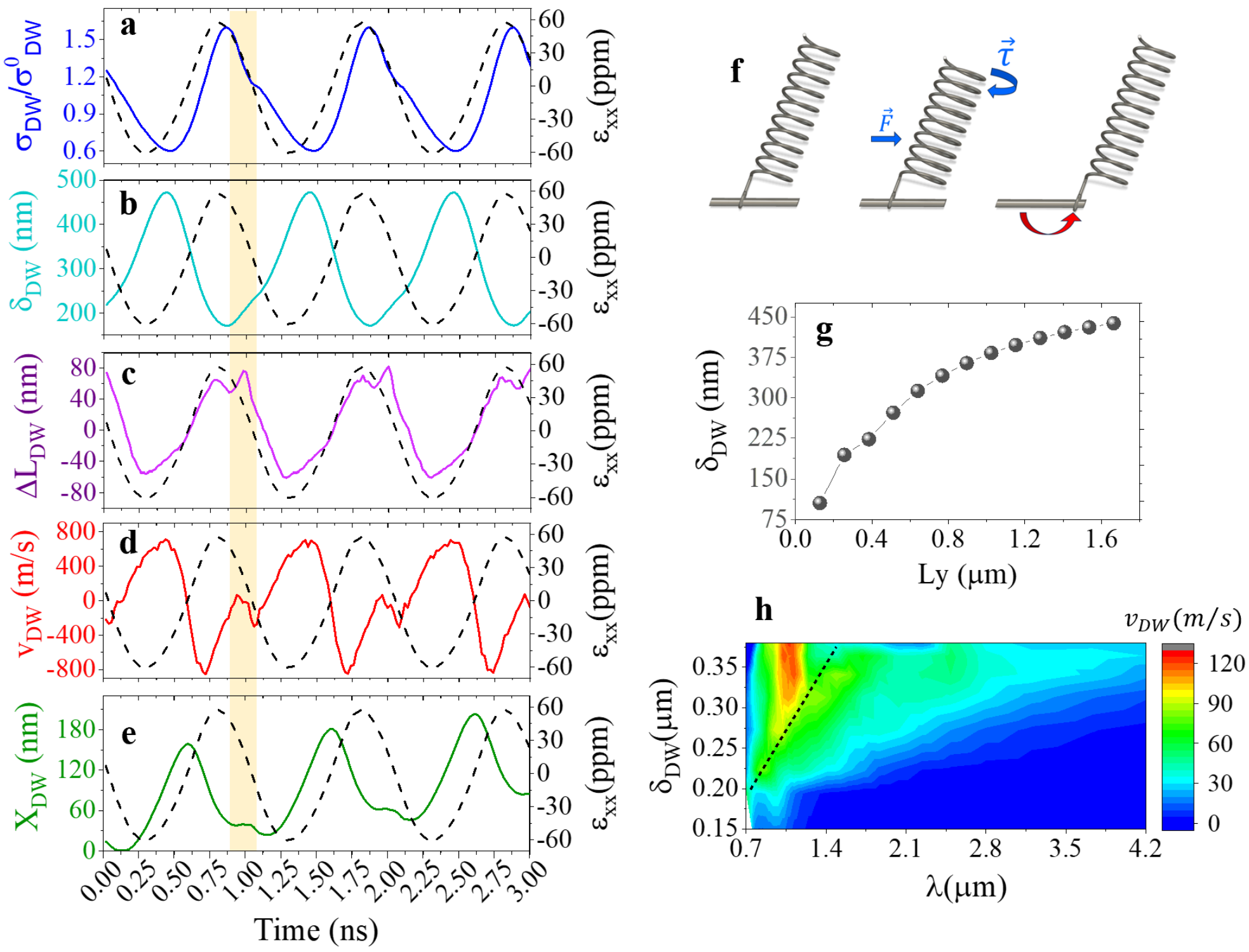

**Fig. 4.** Mechanism of acoustic movement of a magnetic domain wall. (a) Time evolution of the energy of the DW normalized by its resting energy. The dotted black curve represents (in this panel and the subsequent panels) the stress of the SAW evaluated at the center of the moving DW. (b) Time evolution of the width of the DW. (c) Time evolution of the length of the DW. (d) The instant velocity of the DW with time. (e) Position of the DW as a function of time. (f) The mechanical counterpart of the mechanism is described in this article. (g) Width of the DW as a function of the width of the strip. (h) Contour plot with the average velocity of the DW as a function of the width of the DW and the wavelength of the SAW. All the simulations were done for an amplitude of the SAW of $\varepsilon = 70$ ppm propagating from left to right (SAW+). The dotted black straight line in the contour plot has a slope of $\lambda/4$. The relation between SAW frequency and wavelength is given by the velocity of elastic waves in FeCo, $v = \lambda f = 3600$ m/s.

Figure 4f shows a mechanical counterpart of this effect. A tilted spring is twisted while a small force pushes it towards the positive $x$-axis (middle image). If the accumulated stress is large enough, the spring will jump forward to release part of its elastic energy (indicated with a red arrow on the left image).

Therefore, if the right conditions are met, as discussed in the remaining part of this section, the DW moves by 'springing' forward in every cycle of the SAW, gaining linear momentum directly from the SAW. To make the rest of the discussion more readable, we will refer to this complex mechanism as 'DW springing'. One can intuitively realize that, for a given amplitude of the SAW, the DW may be able to spring forward if the DW is wide enough in comparison to the wavelength of the SAW. To determine the importance of the size of the DW, we conducted a systematic micromagnetic simulation for different wavelengths of the SAW and several widths of the strip $L_y$. It is important to note that the DW width $\delta_{DW}$, does not grow linearly with the strip width $L_y$, as shown in Figure 4g. In Figure 4h, we plot the average velocity of the DW as a function of the width of the DW $\delta_{DW}$ and the wavelength of the SAW. The maximum velocity is obtained for a wavelength of $\lambda \approx 1\ \mu m$, for all the strip widths evaluated. Also, the SAW-induced DW motion gets suppressed for $\delta_{DW} \lesssim 0.2\ \mu m$. The dotted straight line in Fig. 4h represents a line with slope $\lambda/4$. It becomes clear that, for the DW to spring under the action of the SAW, its width cannot be far from $1/4$ of the wavelength of the SAW. If the width of the DW is smaller or larger, the movement by springing rapidly becomes less effective or is completely suppressed. This likely accounts for the experimental finding discussed in Fig. 2, where the SAW-induced DW movement is more efficient in strips around 1 μm wide. Interestingly, even for very narrow strips (and narrow DWs), if the wavelength of the SAW is small enough, the DW can still move by springing, even though the average velocity may not be relevant for moderate amplitudes of the SAW, as shown in Suppl. Inf. S8. In any case, Fig. S15 shows that, to achieve a large velocity of the DW under the action of the SAW, it is more effective to adjust the wavelength of the SAW than to increase its power.

It is also important to note that the DW must be asymmetric to spring under the action of the SAW. The effect of the SAW in a vortex DW can be seen in Movie 4 of Suppl. Mat. As the stress accumulated in each cycle of the SAW is symmetric across the DW, the DW only undergoes an oscillation around its position. For a large amplitude of the SAW, the DW moves a small distance (smaller than its width) after many SAW cycles. This is caused by the delay between the accumulation of magnetoelastic stress in the DW and the stress carried by the SAW[19], which may produce a small movement on vortex walls or Skyrmions, but it is too small for any practical purpose.

In Suppl. Info. S9, we compare the SAW-driven DW springing with the movement of the same DW driven by Spin Transfer Torque (STT). Fig. S16 shows that, for any realistic polarization of the current, STT can only achieve DW velocities in the range of 100 m/s if the current density is larger than $10^{12}$ A/m$^2$, which would likely damage the strip irreversibly due to Joule Heating[26]. Interestingly, as explained in Suppl. Info. S9, the DW movement by springing also seems less sensitive to pinning due to grain boundaries than in the case of movement by STT.

To conclude, we have identified a mechanism based on DW compression, through which a SAW can directly transfer linear momentum to an asymmetric transversal DW, and move the DW at sizable velocities in the direction of the SAW propagation. This opens the path for the elusive unassisted and directional control of spin textures by SAW alone, without any external field or electric current. This SAW-induced DW springing could be the key ingredient in many future spintronic devices.

## Methods

**Synchrotron measurements**.

The XMCD-PEEM experiments were performed at the CIRCE beamline of the ALBA synchrotron light source[27] with an Elmitec spectroscopic low-energy electron and photoemission electron microscope (SpeLEEM/PEEM) operating in ultra-high vacuum. The samples were mounted on an in-house designed sample holder[28], and the IDT was contacted with wire bonds. IDTs were several mm away from the sample center and the ferromagnetic strips to screen them from the high electric field of the objective lens by the raised sample holder cap. After introduction in vacuum, samples were degassed at low temperature (<100 °C) for at least 1 h to reduce the risk of arcs between the sample (at high voltage) and the microscope objective (grounded). Also, an acceleration voltage of 10 kV (standard is 20 kV) was used to reduce the risk of discharges. The beamline intensity was adjusted to avoid excessive surface charging of the LiNbO3 substrate. The XMCD-PEEM images are generated from two PEEM images acquired at the energy of the Fe L3 absorption edge with opposite photon helicity (circular positive and negative polarization), and then subtracted pixel-by-pixel to provide the images with XMCD magnetic contrast as intensity. A Keysight signal generator at 650 MHz was used to excite the SAW, followed by a custom-made fiber optical link into the PEEM HV rack[28], where the signal is frequency multiplied to 1.3 GHz and amplified. A dedicated HF cable then connects the HV rack to the sample stage[29].

**Micromagnetic Simulations**.

In the micromagnetic simulations, we assume that the SAW is a pure Rayleigh wave, which consists of both longitudinal ($\sigma_{xx}$ and $\sigma_{zz}$) and shear stress ($\sigma_{xz}$) propagating waves, such as those described in [19]. The stress waves are linked to the corresponding strain waves ($\varepsilon_{xx}$, $\varepsilon_{zz}$ and $\varepsilon_{xz}$) through the elastic constants. In addition, we assume that the strain in the SAW is transferred entirely to the magnetic nanostrip and that the Rayleigh waves propagate without dissipation while traveling through it. Also, magnetostriction, the mechanical strain induced by the magnetic configuration, is neglected. Therefore, the acoustic excitation propagating along the $x$ axis can be as

$$\varepsilon_{xx}(x,t) = E\sin(kx-\omega t)$$

$$\varepsilon_{zz}(x,t) = -E\sin(kx-\omega t)$$

$$\varepsilon_{xz}(x,t) = E'\sin(kx-\omega t+\pi/2)$$

$$\varepsilon_{yy}(x,t) = \varepsilon_{yx}(x,t) = \varepsilon_{yz}(x,t) = 0$$

where $E, \omega$ and $k$ are the SAW amplitude, frequency, and wave number, respectively. In this work $f = 1.3$ GHz and $\lambda = \frac{2\pi}{k} = 2.8$ μm except in Fig. 4h, where both $\omega$ and $\lambda$ are changed according to the dispersion relation $v = \lambda f = 3600$ m/s. Propagation along the positive (negative) x-direction corresponds to $k > 0$ ($k < 0$). The subscript $x$, $y$, and $z$ refer to the coordinate axes in Fig. 3a. The effect of SAW on the magnetization is taken care of via the magnetoelastic field

$$\mathbf{H}_{mel} = -\frac{2}{\mu_0 M_S}\begin{bmatrix} B_1 m_x \varepsilon_{xx} + B_2(m_y \varepsilon_{xy} + m_z \varepsilon_{xz}) \\ B_1 m_y \varepsilon_{yy} + B_2(m_x \varepsilon_{xy} + m_z \varepsilon_{yz}) \\ B_1 m_z \varepsilon_{zz} + B_2(m_x \varepsilon_{xz} + m_y \varepsilon_{yz}) \end{bmatrix}$$

where $B_1$ and $B_2$ are the magneto-elastic constants, and $\mathbf{m} = \mathbf{M}/M_S$ is the normalized magnetization.

The following material parameters were used for FeCo and Py, respectively: exchange constant $A_{\mathrm{FeCo}} = 10.5$ pJ/m, $A_{\mathrm{Py}} = 13$ pJ/m; saturation magnetization $M_{\mathrm{s,FeCo}} = 1.5$ MA/m, $M_{\mathrm{s,Py}} = 0.8$ MA/m; in-plane anisotropy constant $K_{\mathrm{FeCo}} = 7\ \mathrm{kJ/m^3}$, $K_{\mathrm{Py}} = 0\ \mathrm{kJ/m^3}$ (easy axis along $x$ direction) and Gilbert damping constant $\alpha_{\mathrm{FeCo}} = 0.02$, $\alpha_{\mathrm{Py}} = 0.02$. The same magnetoelastic constant values $B_1 = B_2 = -84$ MPa were used for both FeCo and Py and we assumed that the strain induced by the SAW in the PZ is fully transmitted to both magnetic layers. The nanostrip is discretized $4 \times 4 \times 3\ \mathrm{nm^3}$ cells. The grain structure is set using the Voronoi tessellation implemented in mumax3. The average grain diameter is $d = 20$ nm and both the in-plane anisotropy constant and easy-axis orientation are randomly distributed among the different grains following a Gaussian distribution around their nominal values with standard deviation $\sigma = 3\%$.

**Data Availability**

The data supporting the findings of this study are available in the Source data file provided with this paper.

**Acknowledgments**

This work has received funding from the Spanish Ministry of Science and Innovation through projects PID2020-117024GB-C4, PID2020-117024GB-C41, PID2021-122980OB-C53, PID2024-155385NA-C32 and 2D-SAWNICS (PID2020-120433GB-I00). MWK and SRG acknowledge funding through Advanced Materials programme supported by MCIN with funding from the European Union NextGenerationEU (PRTR-C17.I1) and by the Generalitat de Catalunya. MF and MANO acknowledge funding through MCIN by grant PID2021-122980OB-C54.We acknowledge the support from ICTS-MICRONANOFABS through the ICTS-ISOM.

**Author Contributions**

A.R. did or was heavily involved in most of the experimental work, including preliminary electric characterization of the devices, which is not included in the manuscript in its present form. RY did a relevant portion of the micromagnetic simulations with L.L-D and L.T. The Synchrotron experiments were performed by A.R., M.A., J.G and J.L.P., together with the local ALBA team M.F., S.R-G, M.W.K., M.A.N and L.F-G. The IDTs were designed, fabricated and characterized by R.I-L., J.P., F.C. and A.R. Some fabrication techniques were optimized by M.A. The research of A.R. was directed by M.M. and M.M.S. Preliminary Kerr characterization of the devices was performed by G.O-G and S.V. The MFM images were obtained by M.S., R.G., J.G. and J.L.P. The study with micromagnetic simulations was directed by L.L-D, while the experimental research was directed by J.L.P., who wrote the manuscript with input from the other authors.

**Competing Interest**

The authors declare no competing interests.